\documentclass[aps,preprint]{revtex4-1}

\usepackage{epsfig}
\usepackage{latexsym}
\usepackage{amsmath}
\usepackage{amssymb}
\usepackage{amsfonts}
\usepackage{graphicx}
\usepackage{wrapfig}
\usepackage[left]{lineno}
\usepackage{url}
\usepackage{color,bm}

\usepackage[normalem]{ulem}
\usepackage{color}

\definecolor{red}{rgb}{1,0,0}
\definecolor{green}{rgb}{0,1,0}
\definecolor{blue}{rgb}{0,0,1}

\begin{document}

\title{Bistability in orbital trajectories of a chiral self-propelled particle interacting with an external field}

\author{G. A. Patterson}
\email{gpatters@itba.edu.ar}
\affiliation{Instituto Tecnol\'ogico de Buenos Aires, CONICET, Lavard\'en 315, 1437 C. A. de Buenos Aires, Argentina}%

\date{\today}

\begin{abstract}
In this work, the dynamics of a self-propelled stochastic particle under the influence of an axisymmetric light field was experimentally studied. The particle under consideration has the main characteristic of carrying a light sensor in an eccentric location. For the chosen experimental conditions, the emerging trajectories were orbital, and, more interestingly, they presented two preferential radial distances. A mathematical model incorporating the key experimental components was introduced. By means of numerical simulations and theoretical analysis, it was found that, in addition to the orbiting behavior, the sensor location could produce trapped or diffusive behaviors. Furthermore, the study revealed that stochastic perturbation and the eccentric location of the sensor are responsible for inducing bistability in the orbital trajectories, in agreement with the experimental observations.

\end{abstract}

\maketitle

Active matter describes systems which, being out of thermal equilibrium, are composed of agents that consume energy and have the characteristic of being self-propelled \cite{ramaswamy2010mechanics,fodor2018statistical}. These types of systems can have both natural \cite{sokolov2012physical,chen2017weak,xu2019self} and artificial \cite{brambilla2013swarm, hamann2018swarm,dorigo2020reflections,boudet2021collections} origin. Currently, many researchers have focused their attention on studying the interaction of active matter with the environment \cite{bechinger2016active,deblais2018boundaries,dauchot2019dynamics,coduti2019chemotaxis,wang2021emergent,zion2021distributed}. This is of importance since, on the one hand, systems found in nature are always confined or affected by external stimuli \cite{bechinger2016active}, and on the other, synthetic systems are designed to work under such conditions \cite{ozin2005dream,bechinger2016active}. In particular, the motion of particles can be affected by local or global gradients of some external influence. Therefore, effects such as chemotaxis \cite{coduti2019chemotaxis,alirezaeizanjani2020chemotaxis,vuijk2021chemotaxis} and phototaxis \cite{lozano2016phototaxis,mijalkob2016engineering,zion2021distributed} may occur. Notwithstanding, the influence of sensor location on the emerging trajectories remains elusive.

\begin{figure}
\includegraphics{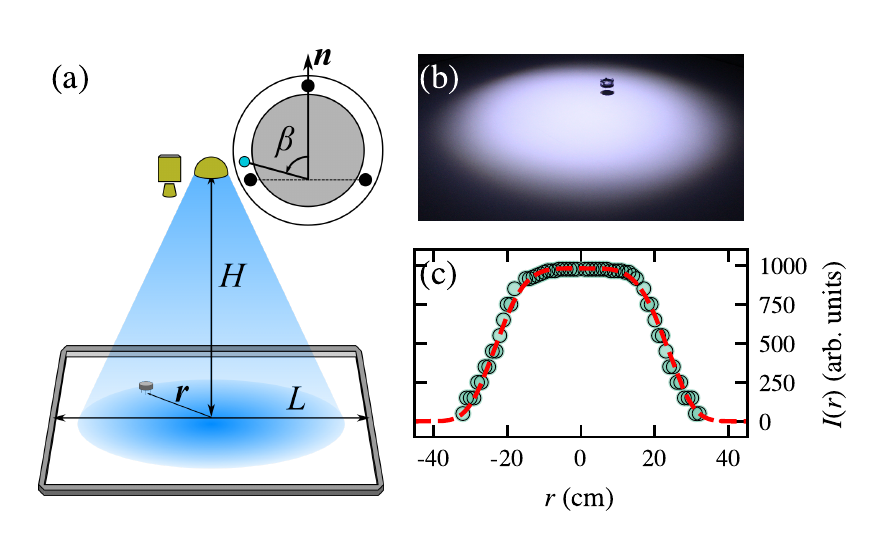}
\caption{\label{fig:expSetup}Experimental setup. (a) The Kilobot is placed over a squared table of length $L=1.2\ \mathrm{m}$, illuminated by a flashlight located at height $H = 1.4\ \mathrm{m}$. Robot top view: the sensor is located at an angle $\beta$ with respect to orientation $\bm{n}$. (b) Photograph of the experiment. (c) Intensity profile measured at the surface of the table. The line stands for the fit of a super-Gaussian function.}
\end{figure}

The main objective of this work focuses on studying the behavior of an isolated active particle that interacts with an external light field through an eccentric sensor. For this aim, a robot commercially available as Kilobot was used \cite{rubenstein2014kilobot}. It features differential locomotion and ambient light sensing capacities, among others. The robot stands on three legs: one front and two rear. Thanks to the differential control of two vibrators, the robot can rotate in one direction or the other around one of its rear legs at a speed of $0.5\ \mathrm{rad/s}$. The Kilobot has the peculiarity that the light sensor is located on top of the robot at an angle $\beta \approx 1.4$ with respect to the orientation of the particle $\bm{n}$ --defined by the front leg and the center of the robot--, as can be seen in Fig.~\ref{fig:expSetup}(a).

The experimental setup consisted of a square table with sides $L = 1.2\ \mathrm{m}$ whose surface was covered by a melamine whiteboard. This table was illuminated by a flashlight located at $H = 1.4\ \mathrm{m}$ in height, projecting an axisymmetric spotlight as illustrated in Figs.~\ref{fig:expSetup}(a)--(b). The experiments were recorded at a rate of 1 fps with a zenith camera located next to the flashlight [see Fig.~\ref{fig:expSetup}(a)]. The propulsion mechanism of the robot was based on consecutive steps caused by turns around one of its rear legs. The counterclockwise turns were around the left leg, while the clockwise turns were around the right leg. Two different step sizes were produced by varying the turning time $T$, that is, the duty cycle of the vibrators. Between each step, the robot stopped for $0.5\ \mathrm{s}$ and measured the light intensity.

The interaction with the light field was based on an algorithm seeking maximum intensity. Two types of interactions were considered: one deterministic and the other stochastic. The first consisted of reversing the turning direction if the intensity value decreased with respect to the measure of the previous step. For the second interaction, the robot made a random choice regarding the direction of the subsequent turn if the intensity of light measured between steps decreased. In both cases, the turning direction remained the same if the intensity value was greater than that of the previous step. Therefore, the interplay between the sensor asymmetry and the interaction with the external field effectively converted the Kilobot into a chiral particle. Studies in systems composed of chiral particles have shown rich emergent behaviors \cite{lowen2016chirality,barois2020sorting,bowick2022symmetry}.

Before carrying out the experiments, the intensity profile was characterized. For this, a Kilobot was programmed to emit a flashing signal with a duty cycle proportional to the intensity received. Figure~\ref{fig:expSetup}(c) shows the radial profile measured from the center of the table. It can be seen to follow a bell-like curve that extends up to approximately $30\ \mathrm{cm}$ from the center.

\begin{figure}
\includegraphics{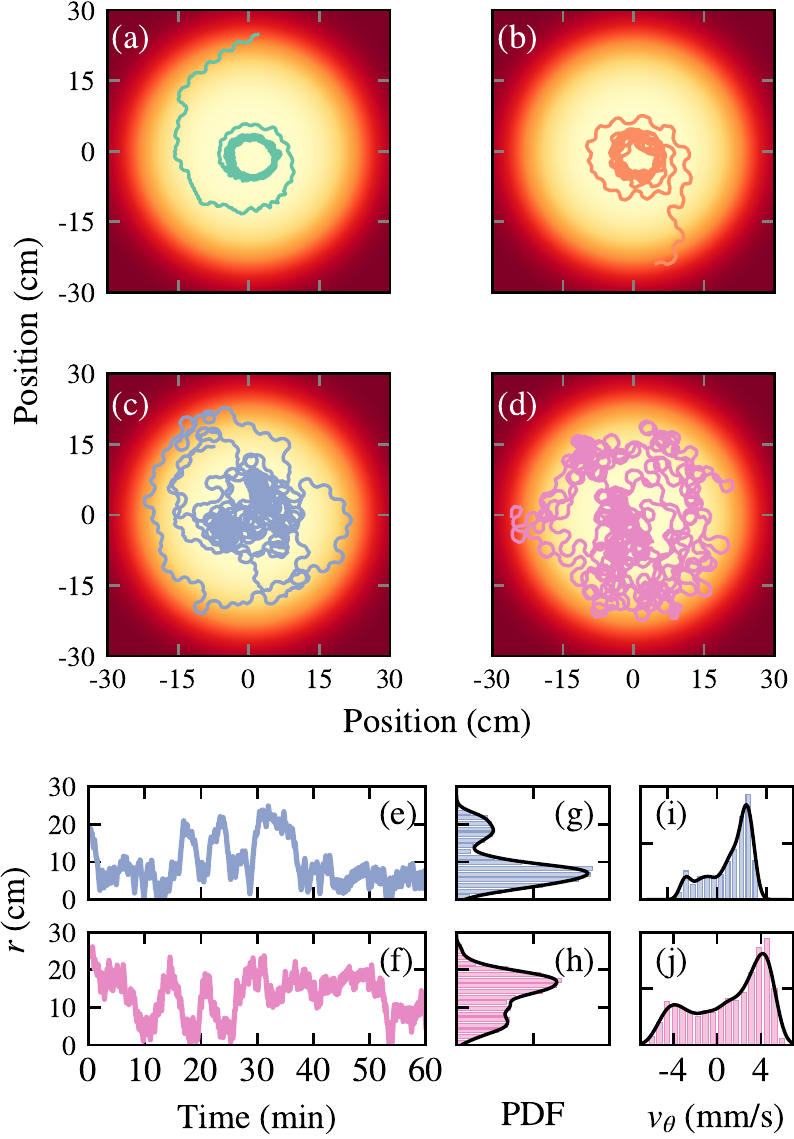}
\caption{\label{fig:expResults1} Experimental results (a)--(b) Deterministic behavior with $T = 1\ \mathrm{s}$ and $T = 2\ \mathrm{s}$, respectively. The Kilobot orbits around the spotlight with the sensor facing towards the center of the field. (c)--(d) Stochastic behavior with $T = 1\ \mathrm{s}$ and $T = 2\ \mathrm{s}$, respectively. The Kilobot performs more complex trajectories but is still biased by the position of the sensor. (e)--(f) Radial position as a function of time for the stochastic behavior with $T = 1\ \mathrm{s}$ and $T = 2\ \mathrm{s}$, respectively. Probability distribution (PDF) of the radial position (g)--(h) and the azimuthal velocity (i)--(j). The black lines stand for the kernel density estimate.}
\end{figure}

Firstly, the deterministic interaction was studied. The Kilobot was placed at an arbitrary position away from the center of the spotlight (between $20$ and $30\ \mathrm{cm}$) and allowed to circulate freely for 15 min. Figures~\ref{fig:expResults1}(a)--(b) show the trajectories made by the robot for $T = 1\ \mathrm{s}$ and $T = 2\ \mathrm{s}$, respectively. It can be seen that, in both cases, the trajectories converge to approximately circular orbits and their direction of circulation is counterclockwise (the sensor always pointing towards the center of the table). These results are in agreement with those found in different types of chiral particles \cite{lowen2016chirality,levis2019activity,barois2020sorting}.

For the experiments involving stochastic interaction, trajectories of $60\ \mathrm{min}$ duration were recorded. Figures~\ref{fig:expResults1}(c)--(d) show that the Kilobot roamed over a larger area, performing more complex trajectories. In the case of $T=1\ \mathrm{s}$, the trajectory followed shorter radial distances, while for $T=2\ \mathrm{s}$, the Kilobot was mostly at farther radial distances. Figures~\ref{fig:expResults1}(e)--(h) show the radial distances as a function of time and their corresponding probability density functions (PDFs). Bimodal distributions reveal that the trajectories present two preferential distances. While for $T=1\ \mathrm{s}$, the PDF has a maximum at $r\approx 5\ \mathrm{cm}$, when turning time is increased to $T=2\ \mathrm{s}$, the maximum value rises to $r\approx 15\ \mathrm{cm}$. Despite its stochastic behavior, the robot follows a  counterclockwise orbital trajectory as shown by the azimuthal velocity $v_\theta$ in Figs.~\ref{fig:expResults1}(i)--(j).

The experimental results suggest that sensor location affects the emergent trajectories by imposing a preferential direction of rotation and, in addition, the stochastic component introduces a bistability effect. To further study these behaviors, a model was designed to include the following factors: (i) alignment between sensor orientation and external field, and (ii) a correlated noise term to account for stochastic steps. The evolution of the position $\bm{r}$ and the orientation $\bm{n} = (\cos{\alpha},\sin{\alpha})$ are
\begin{align}
\dot{\bm{r}} & = v_0\ \bm{n}\ , \\
\tau_0\dot{\alpha}\ \hat{z}& = \bm{s} \times \bm{\nabla}_r I(r)  + \xi\ \hat{z}\ ,
\end{align}
where $v_0$ is the constant velocity of the particle, $\alpha$ is the orientation angle of the particle $\bm{n}$, $\bm{s}= (\cos{\alpha+\beta},\sin{\alpha+\beta})$ is the orientation of the sensor, $\tau_0$ is a characteristic relaxation time, $I(r)$ the field intensity at distance $r$, and $\xi$ Gaussian colored noise with correlation $\langle \xi(t)\xi(t') \rangle = \frac{2D}{\tau_\xi}\exp{-\vert \frac{t-t'}{\tau_\xi}\vert}$ that models the stochastic steps. The light intensity profile was modeled by a super-Gaussian function $I(r)=I_0\exp{\left(- \frac{r}{l_0} \right)}^{a}$ whose parameters were extracted from the data presented in Fig.~\ref{fig:expSetup}(c). The calibration leads to $I_0 = 900$, $l_0 = 250\ \mathrm{cm}$, and $a=4.5$. Rescaling the length by $l_0$ and the time by $\frac{l_0}{v_0}$, the dimensionless equations are
\begin{align}
\label{eq:modelr} \dot{\bm{r}} & = \bm{n}\ , \\
\label{eq:modela} \tau_\alpha\dot{\alpha}\ \hat{z} & = \bm{s} \times \bm{\nabla}_r i(r) + \eta\ \hat{z}\ ,
\end{align}
with $\tau_\alpha = \frac{ v_0}{I_0}\tau_0$, $\eta = \frac{l_0}{I_0}\xi$, $\langle \eta(t)\eta(t') \rangle = \frac{2d}{\tau_\eta}\exp{-\vert \frac{t-t'}{\tau_\eta}\vert}$, $d=\frac{l_0^2}{I_0^2}D$, and $\tau_\eta = \frac{ v_0}{l_0}\tau_\xi$.

\begin{figure}
\includegraphics{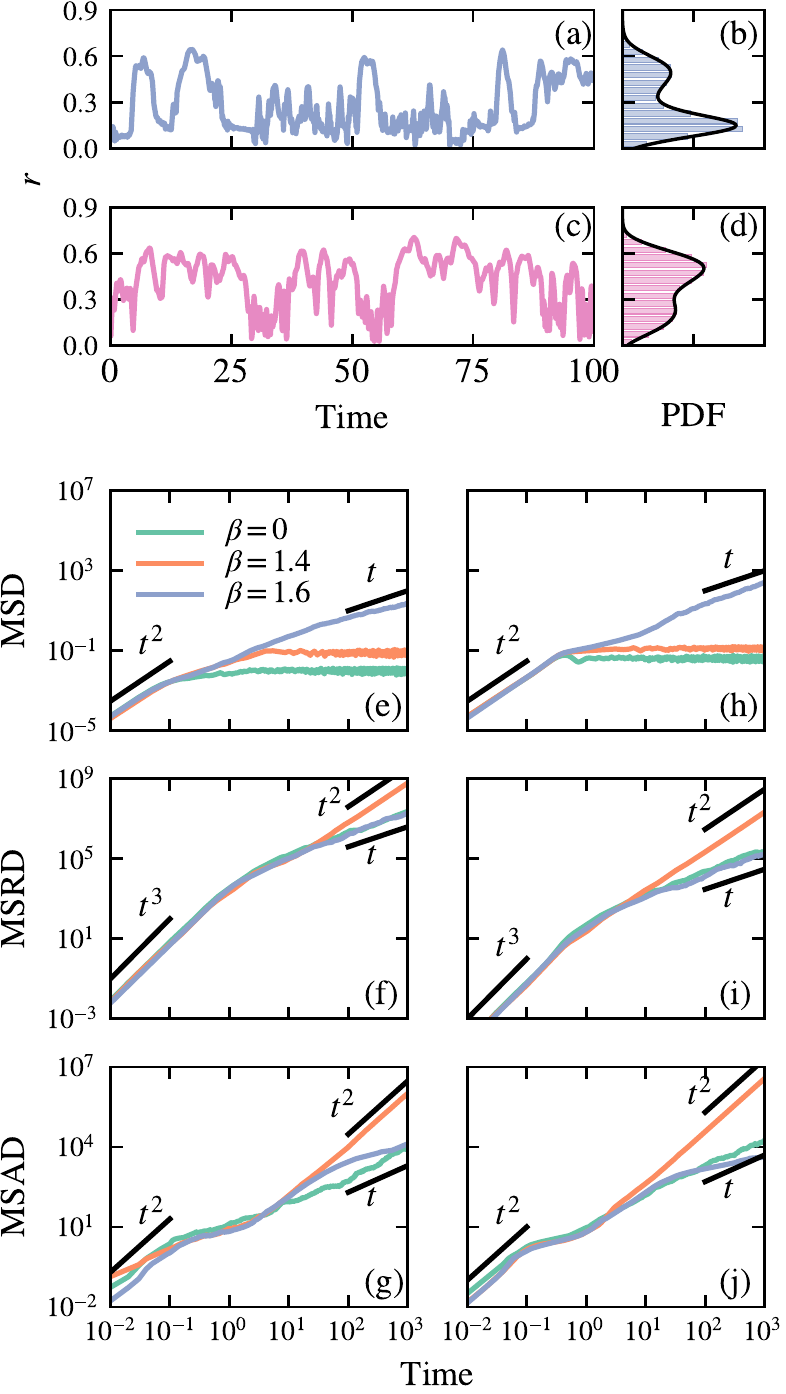}
\caption{\label{fig:simResults} Numerical results. (a)--(d) Radial position as a function of time and the corresponding probability distribution (PDF). The black lines stand for the kernel density estimate. The parameters are: $\tau_\eta=1$, $a=4.5$, $\beta=1.4$, $d=10^{-2}$, and (a)--(b) $\tau_\alpha=10^{-3}$ and (c)--(d) $\tau_\alpha=10^{-2}$. (e)--(j) translational (MSD), rotational (MSRD), and azimuthal mean square displacement (MSAD) for different $\beta$ values. The parameters are: $\tau_\eta=1$, $a=4.5$, $d=10^{-2}$, and (e)--(g) $\tau_\alpha=10^{-3}$ and (h)--(j) $\tau_\alpha=10^{-2}$.}
\end{figure}

The equations were simulated using the Euler-Maruyama algorithm with an integration time step of $dt = 10^{-5}$ and a simulation length of $10^3$. The colored-noise samples were generated from an Ornstein-Uhlenbeck process. Figures~\ref{fig:simResults}(a) and (c) show the radial position $r$ as a function of 100 units of time for $\tau_\alpha = 10^{-3}$ and $\tau_\alpha = 10^{-2}$, respectively. Figures~\ref{fig:simResults}(b) and (d) present the PDFs of the full simulation and show that the model reproduces the bistability behavior of the Kilobot. Remarkably, the effect is independent of the nature of the propulsion mechanism, which is discrete for the Kilobot and continuous in the model. As was demonstrated in the experiments, it is possible to adjust the behavior of the trajectories based on the system's time constants: a low time constant makes the particle move through nearby regions, while a high value favors particle movement through outlying regions.

Next, the role of the sensor location was investigated. For this, considering three values of $\beta$, the mean squared translational ($\mathrm{MSD} = \left\langle \vert \bm{r}-\bm{r}_0 \vert^2 \right\rangle$), rotational ($\mathrm{MSRD} = \left\langle \vert \alpha-\alpha_0 \vert^2 \right\rangle$), and azimuthal ($\mathrm{MSAD} = \left\langle \vert \theta-\theta_0 \vert^2 \right\rangle$) displacements were computed. Regardless of $\tau_\alpha$, Figs.~\ref{fig:simResults}(e)--(j) show that the long-term results can be classified into three different behaviors: (i) for $\beta=0$, the MSD is bounded, and the MSRD and MSAD show diffusive behavior in agreement with the behavior of an active particle trapped within an axisymmetric potential \cite{volpe2013simulation,das2018confined}; (ii) for $\beta=1.4$, the MSD is bounded, and the MSRD and MSAD are ballistic, indicating that the particle orbits around the field center; and (iii) for $\beta = 1.6$, the MSD and MSRD are diffusive, and the MSAD shows subdiffusive behavior, compatible with the diffusive behavior of an isolated particle driven by a dichotomous force \cite{weber2012active}.

Qualitatively, the results for the two $\tau_\alpha$ are similar. The quantitative differences are observed in the crossover times in which the slopes change. In short times, all observables have the same behavior independent of the value of $\tau_\alpha$ and $\beta$: ballistic for the MSD and MSAD, superballistic for the MSRD.

To analyze the stationary behavior of the particle, Eqs.~\eqref{eq:modelr} and \eqref{eq:modela} can be rewritten in polar coordinates as
\begin{align}
\label{eq:r} \dot{r} = & -\sin{\gamma}\ , \\
\dot{\theta} = & \frac{1}{r}\cos{\gamma}\ , \\
\nonumber \dot{\gamma} = & -\left(\frac{1}{r}+\frac{\partial_r i(r)}{\tau_\alpha}\cos{\beta}\right)\cos{\gamma}\\
\label{eq:gamma} & + {\left( \frac{\partial_r i(r)}{\tau_\alpha} \sin{\beta} \right)} \sin{\gamma} + \eta \ ,
\end{align}
where $\gamma$ is the angle formed by the directions $\mathbf{n}$ and $\hat{\theta}$ and is defined as $\gamma = \alpha-\theta -\pi/2$. The steady-state solution of the noiseless system corresponds to an orbital motion if it holds that $\dot{r}=0$, $\ddot{\theta} = 0$, and $\dot{\gamma} = 0$, leading to $\gamma = 0$ and
\begin{equation}
\left(\frac{1}{r}+\frac{\partial_r i(r)}{\tau_\alpha}\cos{\beta}\right) = 0\ .
\label{eq:solution}
\end{equation}
Equation~\eqref{eq:solution} gives the equilibrium distance as a function of the system parameters. Considering that $i(r)=\exp{-r^a}$, the equilibrium positions are
\begin{equation}
r_k^{eq} = \left[-W_k\left(-\frac{\tau_\alpha}{a\cos{\beta}}\right)\right]^{1/a}\ ,
\label{eq:eqposition}
\end{equation}
where $W_k(x)$ is the $k$-branch of the Lambert function. The domain of this function implies that the existence of $r_k^{eq}$ is restricted to
\begin{equation}
\tau_\alpha \leq \frac{a\cos{\beta}}{e}\ ,
\label{eq:limit}
\end{equation}
with $e$ being Euler's number. While the branch $k=0$ gives the stable equilibrium positions, $k=-1$ gives the unstable ones. Both branches annihilate when the equality of Eq.~\eqref{eq:limit} holds, describing a saddle-node bifurcation. The stable solution accounts for the deterministic behavior of the Kilobot shown in Figs.~\ref{fig:expResults1}(a)--(b). When the sensor is aligned to the direction of motion ($\beta = 0$),  Eqs.~\eqref{eq:r}--\eqref{eq:gamma} do not present a stable point for $\gamma$, and the orbital motion does not occur.

Many systems undergo counterintuitive changes when subjeted to stochastic perturbations. For instance, fluctuations may lead to the emergence of new stable states \cite{dorico2005noise,holehouse2020steady}. As the results shown in Figs.~\ref{fig:expResults1}(e)--(h) and \ref{fig:simResults}(a)--(d) suggest that this system presents bistability when stochastic behavior is introduced, the influence of noise is studied by combining Eqs.~\eqref{eq:r} and \eqref{eq:gamma} while approximating $\gamma \approx 0$ and $\dot{\gamma}\approx 0$. In this way, a stochastic differential equation for $r$ is obtained
\begin{align}
\dot{r}	& = -\frac{\frac{\tau_\alpha}{r}+\partial_r i(r)\cos{\beta}}{\partial_r i(r) \sin{\beta}} + \frac{\tau_\alpha}{\partial_r i(r) \sin{\beta}} \eta \\ 
		& = h(r) + g(r) \eta\ .
\end{align}
To calculate the stationary PDF, the unified colored-noise approximation is used following Refs.~\cite{jung1987dynamical,li1995bistable,maggi2015multidimensional,duan2020unified}. Thus, the stationary PDF is 
\begin{equation}
\label{eq:proba} P_s(r) = N \frac{C(r)}{\sqrt{dg^2(r)}} \exp{\left[\int_0^r{\frac{h(r')C(r')}{dg^2(r')}}dr'\right]}\ ,
\end{equation}
where $N$ is a normalization constant and 
\begin{equation}
C(r) = 1 - \tau_\eta g(r)\frac{\mathrm{d}}{\mathrm{d}r}\left(\frac{h(r)}{g(r)} \right)\ ,
\end{equation}
can be interpreted as a correction factor due to the correlation of the stochastic disturbance $\eta$.

\begin{figure*}
\includegraphics{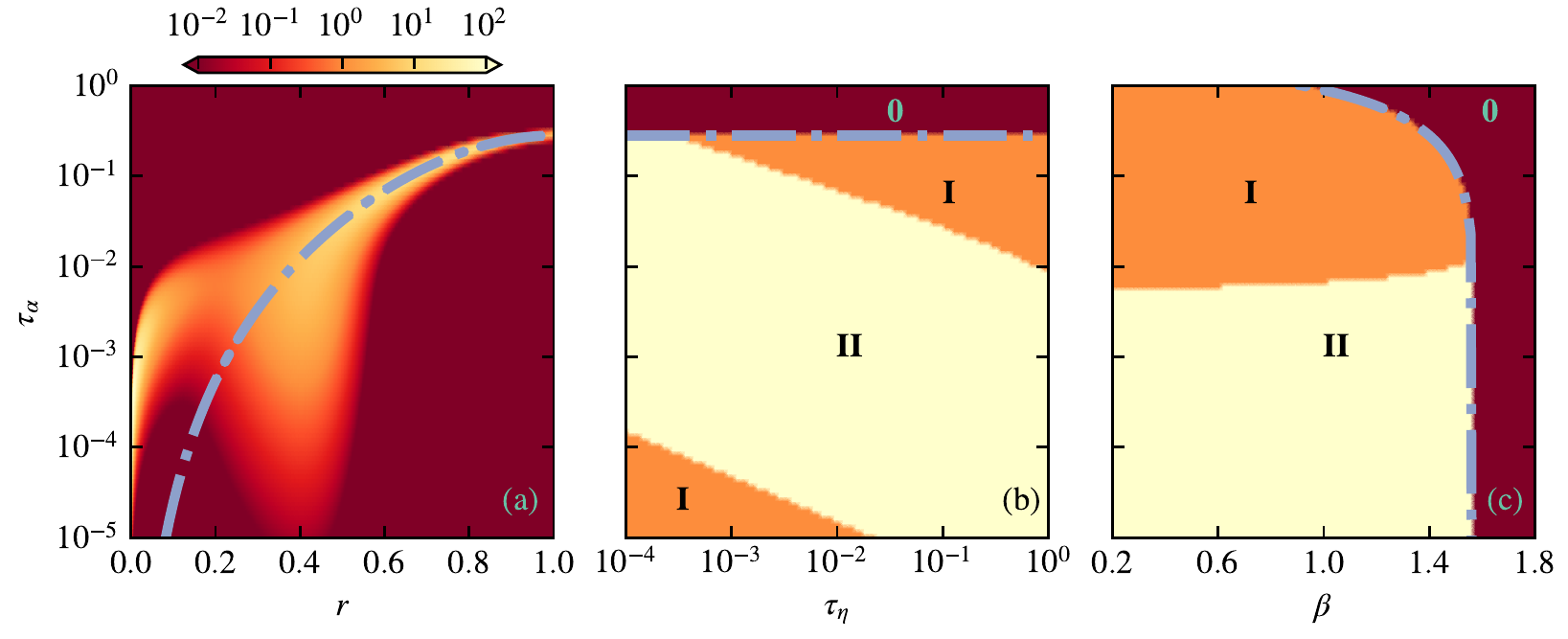}
\caption{Theoretical approximation. \label{fig:theoResults} (a) Stationary probability ($P_s$) as a function of $\tau_\alpha$. The line stands for the deterministic equilibrium position given by Eq.~\eqref{eq:eqposition}. (b)--(c) Stability diagrams as a function of $\tau_\alpha$ and $\tau_\eta$, and $\tau_\alpha$ and $\alpha$, respectively. The solid lines stand for the condition for orbital trajectories given by Eq.~\eqref{eq:limit}. The parameters are: $a=4.5$, $d=10^{-5}$, and (b) $\alpha=1.4$ and (c) $\tau_\eta=1$.}
\end{figure*}

Figure~\ref{fig:theoResults} shows the results obtained from Eq.~\eqref{eq:proba} taking a small value of the perturbation amplitude $d = 10^{-5}$. Figure~\ref{fig:theoResults}(a) shows how $P_s$ varies as a function of $\tau_\alpha$. It can be seen that for values of $\tau_\alpha<10^{-2}$, there is a region of bistability, that is, two local maxima in $P_s$. For higher values, the distribution presents a single maximum that converges to the value of the deterministic solution [Eq.~\eqref{eq:eqposition}]. These results show the dramatic change in the stability of the solutions when considering the stochastic component. In agreement with the experimental results shown in Figs.~\ref{fig:expResults1}(e)--(h), it is observed that within the region of bistability, the absolute maximum of $P_s$ switches position as $\tau_\alpha$ varies. The monostable-bistable transition is also affected by the characteristic time $\tau_\eta$, as shown in the diagram of Fig.~\ref{fig:theoResults}(b). These results were obtained by counting the number of maxima presented by the $P_s$ for each set of parameters. Region \textbf{0} is defined by $P_s$ that do not present any maximum in the interval $r = [0-2]$, regions \textbf{I} are related to monomodal distributions, and region \textbf{II} to bimodal distributions. The black line is the condition given by Eq.~\eqref{eq:limit} that accounts for the orbiting states. The diagram between $\tau_\alpha$ and $\beta$ was also analyzed, as shown in Fig.~\ref{fig:theoResults}(c). Subregions of monostability and bistability can be distinguished within the limit of orbiting states. As can be seen, while the monostable-bistable transition is sensitive to time constants, the position of the sensor has little effect.

In summary, the behavior of a self-propelled particle interacting with an external light field has been studied experimentally. The main characteristic of the particle was that the sensor with which it interacted with the light field was in an eccentric position, turning the Kilobot into a chiral particle. Two interactions based on a maximum intensity searching algorithm were considered: one deterministic and the other stochastic. For the first, it was found that the particle produced orbital paths at a given radial distance, around the intensity maximum. For the stochastic interaction, the emerging trajectories were more complex, and it was found that, in addition to preserving the orbital motion, they presented two preferential radial distances.

To further analyze the observed behavior, a self-propelled particle model was introduced that included a torque term originating from the relative location of the light sensor with respect to the direction of motion. It was found that, depending on the position of the sensor, the trajectories of the particle can be classified into three types: trapped, orbiting, and diffusive.

Finally, to elucidate the emergence of bistability, an analytical approximation of the model was introduced to study the influence of the parameters. It was found that the monostable-bistable transition occurs only in the orbital region and is mostly affected by the characteristic times of the system.

The findings of this work may be of significance to understanding the behavior of biological systems that interact with signals from their environment, as in the design of navigation mechanisms of artificial agents where the sensor location could be changed dynamically to tune different types of trajectories.

\section*{ACKNOWLEDGMENTS}
The author is grateful to D. R. Parisi for his comments and remarks on the manuscript. This work was funded by projects: ITBACyT \#43 (Instituto Tecnol\'ogico de Buenos Aires, Argentina) and PICT 2019-00511 (Agencia Nacional de Promoci\'on Cient\'ifica y Tecnol\'ogica, Argentina).

\bibliography{manusBiblio}

\end{document}